\documentclass[%
 aip,
 amsmath,amssymb,
 reprint,%
]{revtex4-1}

\usepackage{graphicx}
\usepackage{dcolumn}
\usepackage{bm}

\usepackage[utf8]{inputenc}
\usepackage[T1]{fontenc}
\usepackage{mathptmx,amsfonts,amsmath,amsthm,amssymb}

\begin{document}

\preprint{AIP/123-QED}

\title[Thermosize voltage induced in a ballistic graphene nanoribbon junction]{Thermosize voltage induced in a ballistic graphene \\ nanoribbon junction}

\author{Alhun Aydin}
\affiliation{Department of Physics and Astronomy, Box 516, 75120, Uppsala University, Uppsala, Sweden}
\affiliation{Nano Energy Research Group, Energy Institute, Istanbul Technical University, 34469, Istanbul, Turkey}
\author{Jonas Fransson}%
\affiliation{Department of Physics and Astronomy, Box 516, 75120, Uppsala University, Uppsala, Sweden}

\author{Altug Sisman}
 \email{altug.sisman@physics.uu.se}
\affiliation{Department of Physics and Astronomy, Box 516, 75120, Uppsala University, Uppsala, Sweden}
\affiliation{Nano Energy Research Group, Energy Institute, Istanbul Technical University, 34469, Istanbul, Turkey}

\date{\today}

\begin{abstract}
A thermoelectric voltage is induced in a junction, constituted of two dissimilar materials under a temperature gradient. Similarly, a thermosize voltage is expected to be induced in a junction made by the same material but having different sizes, so-called thermosize junction. This is a consequence of dissimilarity in Seebeck coefficients due to differences in classical and/or quantum size effects in the same materials with different sizes. The studies on thermosize effects in literature are mainly based on semi-classical models under relaxation time approximation or even simpler local equilibrium ones where only very general ideas and results have been discussed without considering quantum transport approaches and specific materials. To make more realistic predictions for a possible experimental verification, here, we consider ballistic thermosize junctions made by narrow and wide ($n$-$w$) pristine graphene nanoribbons with perfect armchair edges and calculate the electronic contribution to the thermosize voltage, at room temperature, by using the Landauer formalism. The results show that the maximum thermosize voltage can be achieved for semiconducting nanoribbons and it is about an order of magnitude larger than that of metallic nanoribbons. In the semiconducting case, the thermosize voltage forms a characteristic plateau for a finite range of gating conditions. We demonstrate, through numerical calculations, that the induced thermosize voltage per temperature difference can be in the scale of mV/K, which is high enough for experimental measurements. Owing to their high and persistent thermosize voltage values, graphene nanoribbons are expected to be good candidate for device applications of thermosize effects.
\end{abstract}

\maketitle

\section{Introduction}

Thermoelectric effect provides a unique mechanism for direct conversion between heat and electricity. Pioneering works by Hicks and Dresselhaus in 1993 \cite{te1,te2} stimulated extensive research on enhancing thermoelectric efficiency by making use of quantum confined structures \cite{te3}. Since then, thermoelectric junctions have been explored in various contexts, such as quantum dots \cite{tedot}, molecules \cite{tejunc2007a,tejunc2011a,tejunc2018a}, nanowires \cite{tenw}, superlattices \cite{norvec1,norvec2}, and graphene\cite{tegnr2007a,tegnr2009a,tegnr2012a,tegnr2012b,tegnr2012c,tegnr2013a,tegnr2014a,tebook2014,tegnr2015a,tegnr2017a,tegnr2018a}.
Rather than making a junction between electrodes with different materials, a junction between electrodes with the same material, however, having different sizes, may also generate an electrochemical potential difference when a temperature gradient is applied. This potential difference emerges from distinctive classical and/or quantum size effects on the Seebeck coefficients, so-called thermosize effect  \cite{tse1}. The thermosize effect does not only open new optimization possibilities in thermoelectric junctions, but also serves as a new computational and experimental platform for examination of quantum size effects in different materials. Various studies have been conducted for thermosize effect and their possible applications during the last fifteen years  \cite{tsef2008a,tsef2008b,tsef2010a,tse2,tse3,tsef2011a,tsef2012a,tsef2012b,tse4,tsef2013a,tsef2013b,tsef2014a,tse5,tsef2018a,tsef2018b}.

Thermosize effect is initially proposed and designed for junctions between the electrodes of the same material, in which one electrode is at nanoscale whereas the other one is at macroscale. In this way, the set-up constitutes a nano-macro thermosize junction (TSJ) \cite{tse1}. By considering nano-macro as well as nano-micro junctions, thermodynamic performance analyses of cycles working with both classical and quantum thermosize effects have been examined thoroughly \cite{tsef2008a,tsef2008b,tsef2010a,tse2,tse3,tsef2011a,tsef2012a,tsef2012b,tse4,tsef2013a,tsef2013b,tsef2014a,tse5,tsef2018a}. On the other hand, these studies in literature are based on either semi-classical models using relaxation time approach or even simpler local thermodynamic equilibrium ones. Therefore, some general ideas and results have been discussed without considering quantum transport approaches and specific materials. In addition, despite the fact that quantum size effects are expected to be much more prominent for ballistic transport \cite{bineker}, thermosize effect in quantum ballistic regime has been completely overlooked.

In this study, we demonstrate, through numerical calculations, that the induced thermosize voltage between quantum ballistic electrodes can become high enough to experimentally verify and use it for device applications. Here, we choose armchair graphene nanoribbons (GNRs), having both width and length smaller than the phase coherence length, as well as the characteristic mean free path of electrons, as electrode material in our TSJ. We calculate the transmission function of the GNRs, described using a tight-binding model, in the transmission (Landauer) formalism. From the transmission we obtain the thermoelectric transport coefficients and thermosize potential for GNR TSJs, as function of size and aspect ratio. We examine, using a few viable examples, the thermoelectric properties of GNRs as well as thermosize voltage in GNR TSJs.

GNR is a perfectly suitable material for the TSJ due to its ballistic transport properties at room-temperature \cite{gnr2007a,gnr2014a,gnr2016a}. Moreover, owing to the strong size dependence of the electronic properties and corresponding band structure \cite{RevModPhys.83.407,tegnr2012b}, GNRs serve as one of the best materials to investigate size-dependent properties. They may, in addition, be good candidates for realizations of TSJs, considering the efforts on their practical usage \cite{tebook2014,gnr2016a,tegnr2018a}. Depending on the shape of their edges, GNRs are classified as zig-zag or armchair GNRs. Since the thermoelectric performance of armchair GNRs is shown to be better than for zig-zag GNRs in general  \cite{tegnr2012a}, we consider narrow-wide ($n$-$w$) armchair GNR TSJs in this study. 

\section{Thermosize junction of graphene nanoribbons and its modeling}

\begin{figure}[t] 
\centering
\includegraphics[width=\columnwidth]{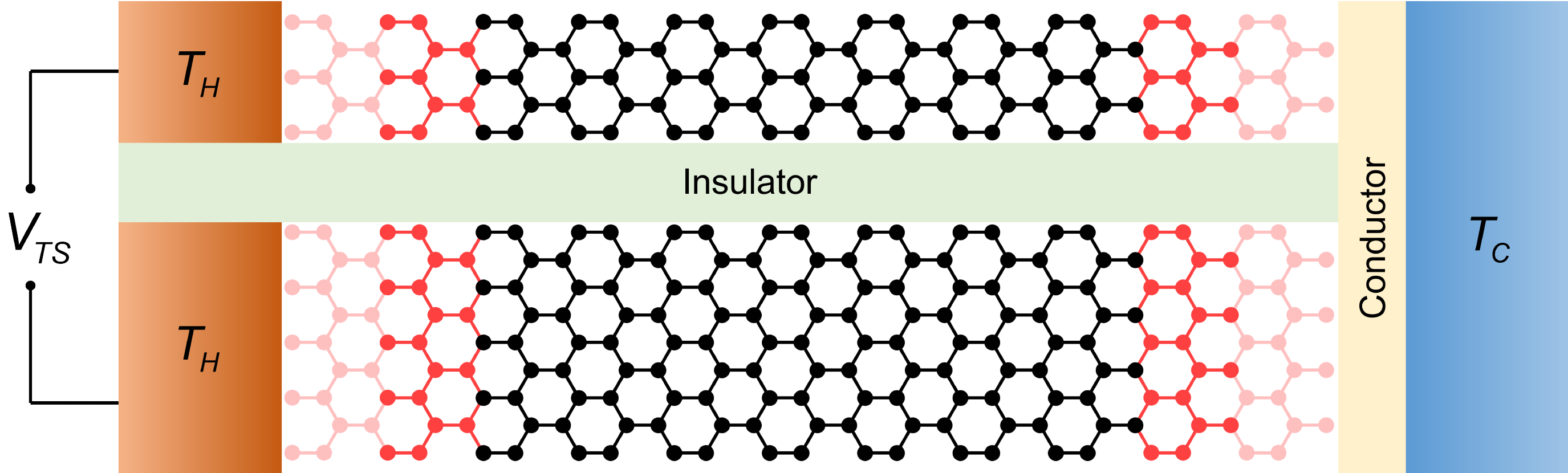}
\caption{A schematic (not to scale) of a thermosize junction (TSJ) of narrow-wide ($n$-$w$) graphene nanoribbons (GNRs) with armchair edges (armchair GNR). Temperature difference $(T_H>T_C)$ at the ends of the same material with different sizes gives rise to thermosize voltage $V_{TS}$.}
\label{fig:pic1}
\end{figure}

A schematic view of the set-up we propose is presented in Fig. \ref{fig:pic1}, showing a TSJ of a narrow and wide ($n$-$w$) armchair GNRs. A temperature gradient, defined between by a set of hot thermal reservoir ($T_H$=310 K) at the left end and a cold thermal reservoir ($T_C$=300 K) at the right. At the cold end of the junction, the two GNRs are interconnected to one another, both electrically and thermally. An insulating layer separating the GNRs from each other extends into the hot reservoir, thereby, electrically separating the GNRs along the transport direction. Under zero external bias voltage, an electrochemical potential difference is induced between the hot and cold reservoirs across the GNRs, emerging as a response to the applied temperature difference. However, because of the unequal sizes (widths) of the two GNRs, the induced electrochemical potential differences across the GNRs are different. Hence, since the electrochemical potentials of the GNRs are equal at the cold side, a voltage difference is generated between the GNRs at the hot end of the junction. This phenomenon is referred to as the thermosize effect. Note that both transport and transverse directions are much smaller than the mean free path and comparable with the de Broglie wavelength of electrons. The width per dimer is around 0.12 nm and depending on the number of dimers, the width of GNRs ranges from 0.6 nm (5 GNRs) to 3 nm (25 GNRs).

In order to quantify our discussion, we construct a mathematical model corresponding to the system in Fig. \ref{fig:pic1}. The GNRs can be effectively described by a tight-binding Hamiltonian,
\begin{equation}
{\cal H}=-t\sum_{\left\langle i,j\right\rangle}c_i^\dagger c_j+H.c.
\end{equation}
where the sum runs over all nearest-neighbors $\left\langle i,j\right\rangle$, $t$ is the hopping rate, the operator $c_i^\dagger$ ($c_i$) creates (annihilates) an electron on the $i$:th site. The on-site energy is chosen to be zero relative to the Fermi energy which we set to $\epsilon_F=0$. For the quantum transport calculations we have employed the software \textsc{KWANT} \cite{kwant}, which allows for construction of the tight-binding model and calculation of total transmission function ${\cal T}(\varepsilon)$. The nearest-neighbor hopping rate $t=2.7$ eV, which successfully describes the electronic properties of GNRs in the absence of lattice deformation\cite{gnr2007a,PhysRevB.66.035412,PhysRevB.81.245402}. Thanks to the weak electron-phonon coupling in GNRs, even at room temperature, phonon scattering of electrons is negligible\cite{gnr2007a,tegnr2009a,tegnr2012b}. We assume that the electrical contacts are perfectly transmitting, which leads to that the thermoelectric coefficients only depend on the widths of GNRs. In this way, we can focus on solely the influence of the size (width) difference.

Depending on its width, the GNR can be either semiconducting or metallic \cite{RevModPhys.83.407}. The width of the GNR is associated with the number $N$ of dimer lines across its width, and we shall use this number to label the N-armchair GNR. The junctions constructed at the interface between semiconducting and metallic GNRs cannot be considered as TSJs, since semiconducting and metallic GNRs are actually different types of materials in this context. Therefore, such a set-up constitute a usual thermoelectric junction, and while size matters also for such junctions we shall omit this possibility since we want to focus on thermosize effects only. Before we approach the construction of TSJs, we first investigate the thermoelectric transport properties like conductance, Seebeck coefficient, and power factor of GNRs as function of the width, in order to understand their individual thermoelectric properties at room temperature for different widths of AGNRs. 

By employing transmission formalism \cite{dattabook1995}, the dimensionless transport integral in the linear response regime reads
\begin{align}
I_\alpha=&
	\int\left[\beta\left(\varepsilon-\mu\right)\right]^\alpha \beta f(\varepsilon)[1-f(\varepsilon)]{\cal T}(\varepsilon) d\varepsilon
	,
\end{align}
where $\beta=1/(k_B T)$, with the Boltzmann constant $k_B$ and temperature $T$, $\alpha$ indicates the energy moment index, $\mu$ denotes the chemical potential, and $f(\varepsilon)=1/\{\exp[\beta(\varepsilon-\mu)]+1\}$ is the Fermi-Dirac distribution function. In terms of the transport integral $I_\alpha$, the electrical conductance and Seebeck coefficient are then written as
\begin{subequations}
\begin{align}
G=&
	\frac{2e^2}{h}I_0
	,
\\
S=&
	-\frac{k_B}{e}\frac{I_1}{I_0}
	,
\end{align}
\end{subequations}
where $e$ is the electron charge, $h$ is the Planck constant, and the factor 2 signifies spin degeneracy. The power factor $P$ is then given by $P=GS^2$.

\begin{figure*}[t]
\centering
\includegraphics[width=0.85\textwidth]{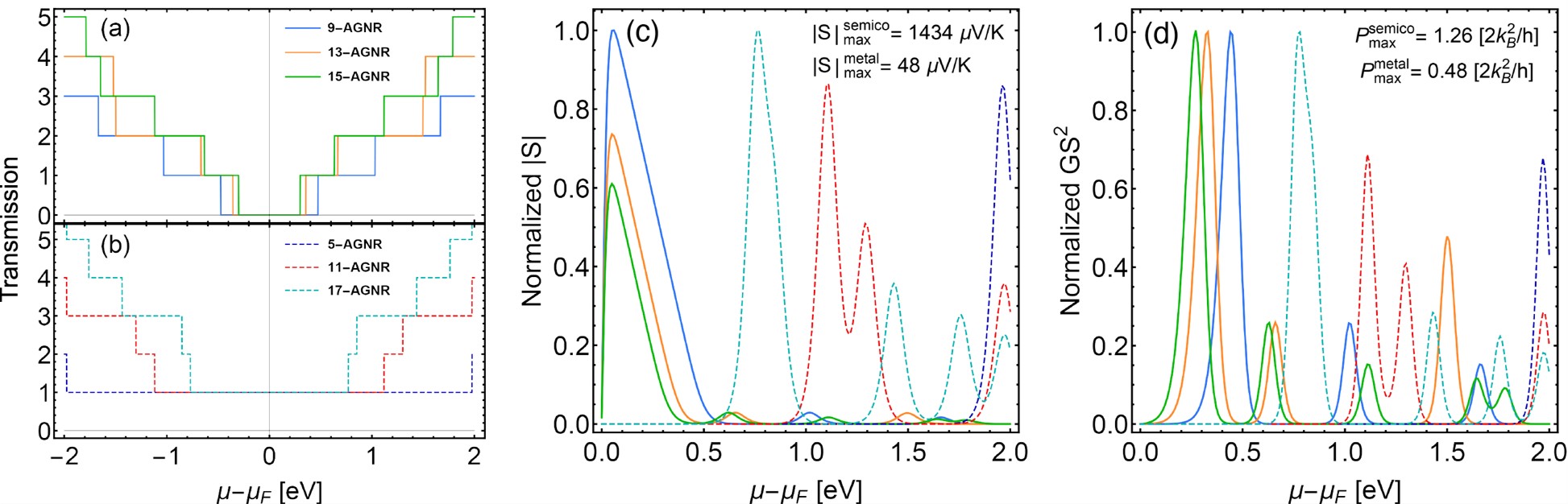}
\caption{The transmission coefficient ${\cal T}$ for (a) semi-conducting and (b) metallic GNRs, (c) Seebeck coefficient and (d) power factor (the last two are normalized to their maximum values) at 300 K as function of the chemical potential for GNRs having 6 different widths, where the 9-, 13- and 15-armchair GNRs exhibit semiconducting, while the 5-, 11- and 17-armchair GNRs show metallic transport properties.}
\label{fig:pic2}
\end{figure*}

In Fig. \ref{fig:pic2} (a) and (b), we plot the transmission coefficients as function of the chemical potential $\mu$ for a few GNRs with different $N$ ranging between 5 and 17. From these plots it can be concluded that we can classify the GNRs into two groups with distinct conducting properties. The first class, comprising the 5-, 11-, 17-armchair GNRs, as well as the 23-armchair GNR (not shown here), all have zero band gap, and would, therefore, be considered as metallic. The second class, containing the 9-, 13-, 15-armchair GNRs, complemented by the 7-, 19-, 21-, and 25-armchair GNRs (not given here), have finite band gaps and are, thereby, here regarded as semiconducting. This latter class is identified by band gaps in the order of more than 400 meV (25-armchair GNR giving the minimum band gap), which, hence, provide viable semiconducting properties at room temperature. Our results are in very good agreement with the previous ones in literature \cite{gnr2016a}.

From the transmission coefficient we extract the Seebeck coefficient and power factor. In Fig. \ref{fig:pic2} we plot (c) the modulus of the Seebeck coefficients and (d) the power factor (both normalized to their corresponding maximum values) as function of the electrochemical potential $\mu$, for three metallic and three semiconducting GNRs. Since we want to focus only on a certain type of carriers in the TSJs, we consider conduction electrons for which $\mu>0$ (n-type). As expected \cite{tegnr2012b,tegnr2014a}, the Seebeck coefficients of the semiconducting GNRs are substantially, more than two orders of magnitude, larger than for the corresponding metallic ones. This is understood to be an effect of the finite band gap around the Dirac point for the semiconducting GNRs. In spite of the huge difference between the maximums of Seebeck coefficients of semiconducting and metallic GNRs, the discrepancy between their corresponding power factors is not comparably large. Naturally, this originates from the (much) larger conductances of the metallic GNRs compared to those of semiconducting ones in the pertinent ranges of chemical potentials. Thus, from the result presented in Fig. \ref{fig:pic2}, it can be concluded that the thermosize voltage should be significant for semiconducting GNRs at low chemical potentials.

\section{Thermosize voltage in graphene nanoribbon junctions}

In a TSJ under zero external bias voltage, the applied temperature difference first serves as the sole driving force for the charge current. Then, because of the zero steady-state current condition, an electrochemical potential difference builds up as an opposite driving force for the current. The net particle current inside each GNR of the TSJ can be expressed as
\begin{align}
I_\text{net}=&
	I_H-I_C
	=
	\frac{2e}{h}\int\left[f\left(\mu_H,T_H\right)-f\left(\mu_C,T_C\right)\right]{\cal T}(\varepsilon)d\varepsilon
	,
\end{align}
under the assumption that ${\cal T}(\varepsilon)$ is the same for left and right moving particles. Here, the hot (cold) end of the junction is denoted by $H$ ($C$). The emergent thermosize voltage between the narrow and wide GNRs is then defined as
\begin{align}
V_{TS}(\mu_C,T_H,T_C)=&
	\mu_H^n\big|_{I_{net}=0}-\mu_H^w\big|_{I_{net}=0}
	.
\end{align}

\begin{figure*}[t]
\centering
\includegraphics[width=0.8\textwidth]{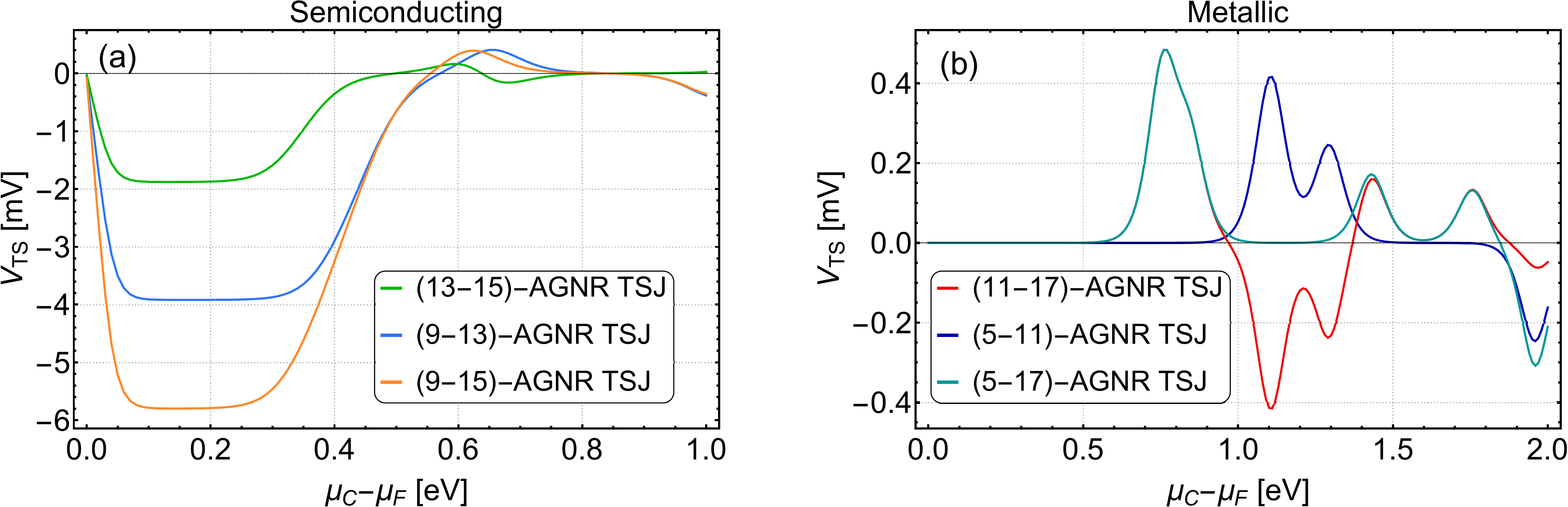}
\caption{Thermosize voltage changes with the chemical potential of cold side for (a) semiconducting and (b) metallic TSJs of armchair GNRs for various ($n$-$w$) width combinations for 10K temperature difference at room temperature.}
\label{fig:pic3}
\end{figure*}

In Fig. \ref{fig:pic3} we plot the thermosize voltages for three different (a) semiconducting and (b) metallic GNR TSJs configurations, as function of the chemical potential $\mu_C$ of the cold side. For semiconducting  TSJs, the magnitude of thermosize voltage grows rapidly with increasing values of $\mu_C$ in the vicinity of zero and then forms an extended plateau for a wide range of $\mu_C$ and returns back to a negligible voltage in an oscillatory fashion for large values of $\mu_C$.
The thermosize voltage in the metallic TSJs, on the other hand, remains vanishingly small for a wide range of $\mu_C$ and becomes finite only at large values of $\mu_C$, acquiring an oscillatory behavior about zero voltage for increasing $\mu_C$.
The results, hence, suggest that for large values of electrochemical potential $\mu_C$, the thermosize voltages generated by semiconducting TSJs, on the one hand, and metallic TSJs, on the other, are comparable in magnitude.
By contrast, for small values of $\mu_C$ the thermosize voltages generated in semiconducting TSJs are about 10 times larger than in the corresponding metallic TSJs.
Comparing the results in Figs. \ref{fig:pic2} (b) and \ref{fig:pic3} (a), we see that thermosize voltage is sizable in the same range of chemical potentials where the Seebeck coefficient peaks. The large magnitude of the thermosize voltage and the characteristic plateau for semiconducting TSJs suggest that this set-up should be promising for the experimental demonstration of the effect.

\begin{figure*}[t]
\centering
\includegraphics[width=0.8\textwidth]{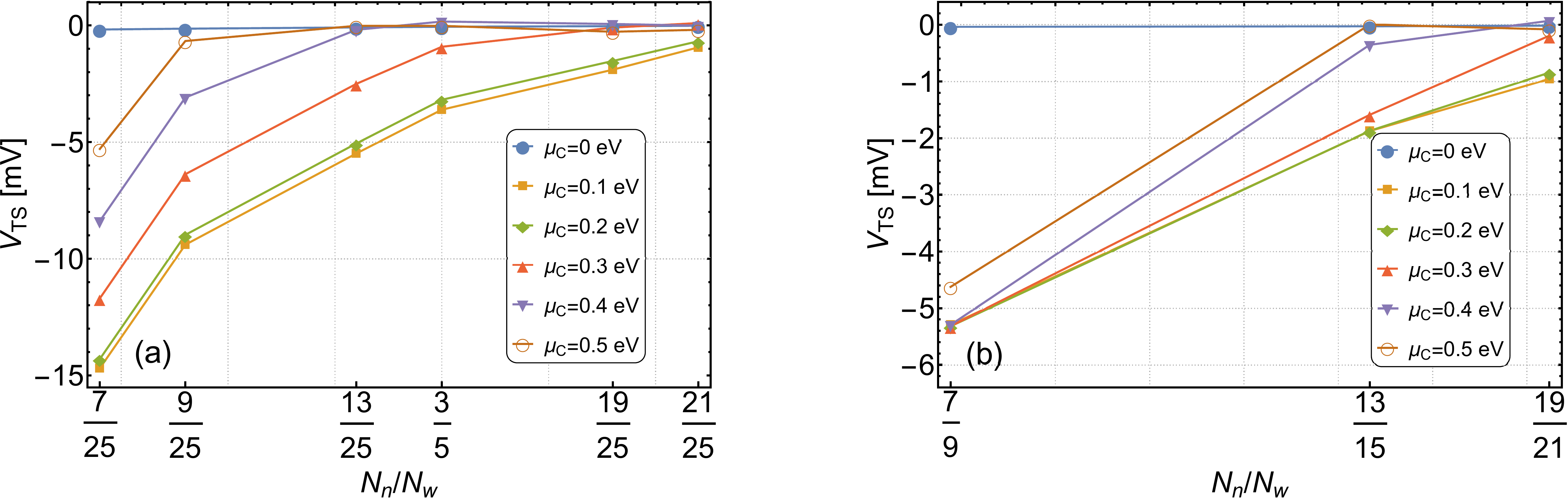}
\caption{Aspect ratio dependences of thermosize voltage in semiconducting TSJs plotted for different chemical potential values (a) by changing narrow side while keeping the wide side constant and (b) by changing both sides while keeping $N_w-N_n=2$.}
\label{fig:pic4}
\end{figure*}

It can also be noticed that the thermosize voltages in metallic TSJs are similar in magnitude for different GNRs configurations. This can be contrasted by the observation that the thermosize voltages for different GNRs configurations in semiconducting TSJs differ by almost a factor of three. In fact, the results in Fig. \ref{fig:pic3} (a) indicate that, a simultaneous increase of the widths of both $n$ and $w$ GNRs, tends to decreases the thermosize voltage. This is an expected behavior since the quantum size effects become decreasingly important for larger sizes. Higher values of the aspect ratio between the widths of $n$ and $w$ GNRs, however, cause higher thermosize voltage. This behavior clearly suggests that the larger the difference in magnitudes of quantum size effects, the larger thermosize voltage can be obtained. In both semiconducting and metallic cases, the sign of the thermosize voltage can be controlled by the chemical potential.

The aspect ratio of $n$-$w$ is important for the design of the TSJ. Since the semiconducting TSJs give much higher thermosize voltages, we restrict our further considerations to this class of junctions. The plots in Fig. \ref{fig:pic4} show the variations of the thermosize voltage as function of the aspect ratio, for different values of $\mu_C$. In Fig. \ref{fig:pic4} (a), we vary the aspect ratio $n/w$, between $7/25\lesssim0.3$ and $21/25\gtrsim0.8$, of the junction by keeping the width of the wide GNR constant. Here, we consider only the configurations of GNRs which leads to semiconducting TSJs. The results clearly shows that the thermosize voltage becomes larger the smaller the aspect ratio is, and it decreases monotonically with increasing aspect ratio. However, since quantum size effects become more prominent for the smaller sizes, we also checked whether a near unity aspect ratio can still sustain a considerable thermosize voltage. In Fig. \ref{fig:pic4} (b) we plot the thermosize voltage as function of aspect ratio in the range between $7/9\gtrsim0.77$ and $19/21\lesssim0.91$ by keeping the difference $N_w-N_n$ between the wide and narrow GNRs constant. These results clearly show that even an aspect ratio near unity may not be detrimental for small enough GNRs constituting the junction. Intrinsic quantum size effects may very well become stronger and make the potential difference high enough even though the widths of GNRs are close to each other. This perspective is also corroborated by the fact that the thermosize voltage monotonically decreases with increasing widths of the GNRs, keeping the difference $N_w-N_n$ fixed. For wider GNRs, the intrinsic quantum size effects become increasingly similar in the two GNRs which, therefore, tends to diminish the electrochemical potential difference between the electrodes.

\section{Conclusion}
In conclusion, we predict that thermosize effect can experimentally be verified and examined by constructing GNR TSJs. In the studied set-ups we predicted voltages per unit temperature difference in the order of mV/K, at room temperature. This suitability of GNRs for TSJs is also accompanied by their preserved ballistic transport properties and low electron-phonon coupling at room temperature, as well as the strong quantum size dependence of their electronic properties and band structures. In this realms, the atomically thin GNR allows us to explore the thermosize effect upon approaching the quantum limit. One shall notice that our results may be regarded as an upper limit for the thermosize voltage, since deviations from these ideal results may arise from, for example, impurities, contact resistances, and incoherences. We nonetheless believe that the effect remains sufficiently large in GNRs due to state-of-the-art capabilities to manufacture ultra-clean graphene \cite{clean1,clean2}. In a realistic device, the GNR most likely have to be deposited on a substrate, which may open up new opportunities in the design of energy conversion devices based on quantum size effects.

\bibliography{tseref}
\bibliographystyle{unsrt}
\end{document}